\documentstyle[aps,12pt,epsf,amssymb]{revtex}
\begin{document}
\baselineskip=0.8cm
\newcommand{\ini}{\begin{equation}}
\newcommand{\fin}{\end{equation}}
\newcommand{\inir}{\begin{eqnarray}}
\newcommand{\finr}{\end{eqnarray}}
\newcommand{\inif}{\begin{figure}}
\newcommand{\finf}{\end{figure}}
\newcommand{\bc}{\begin{center}}
\newcommand{\ec}{\end{center}}
\def\ol{\overline}
\def\pa{\partial}
\def\ra{\rightarrow}
\def\ts{\times}
\def\df{\dotfill}
\def\bs{\backslash}
\def\dg{\dagger}

$~$

\hfill DSF-T-99/46

\vspace{1 cm}

\centerline{\LARGE{Heavy neutrino mass scale}}

\centerline{\LARGE{and quark-lepton symmetry}}

\vspace{1 cm}

\centerline{\large{D. Falcone}}

\vspace{1 cm}

\centerline{Dipartimento di Scienze Fisiche, Universit\`a di Napoli,}
\centerline{Mostra d'Oltremare, Pad. 19, I-80125, Napoli, Italy}

\centerline{{\tt e-mail: falcone@na.infn.it}}

\vspace{1 cm}

\begin{abstract}

\noindent
Assuming hierarchical neutrino masses we calculate the heavy neutrino
mass scale in the seesaw mechanism
from experimental data on oscillations of solar and atmospheric
neutrinos and quark-lepton symmetry. The resulting scale is around or above the
unification scale, unless the two lightest neutrinos
have masses of opposite sign, in which case the resulting scale can
be intermediate.
 


\end{abstract}

\newpage

\noindent
Recent results on atmospheric and solar neutrinos \cite{98} support the idea
that neutrinos have tiny masses. A popular mechanism for achieving small
Majorana masses for left-handed neutrinos is the seesaw mechanism
\cite{seesaw}, where the Majorana mass matrix of the left-handed neutrino is
given by
\ini
M_L=M_D M_R^{-1} M_D^T,
\fin
with $M_D$ the Dirac mass matrix and $M_R$ the Majorana mass matrix of a heavy
right-handed neutrino, related to lepton number violation at high energy
\cite{lep}. If the Dirac masses are supposed to be of the same order
of magnitude of the up-quark masses (quark-lepton symmetry),
as suggested by GUTs \cite{guts}, then light
left-handed neutrinos appear. In such a case the heavy neutrino mass scale
should be at the unification or intermediate scale \cite{lep},
because in most GUTs the Higgs field that gives mass to neutrinos is
the same that breaks the GUT group or the intermediate
group to the standard model. Also, at that scale the relation $m_b=m_{\tau}$,
resulting from quark-lepton symmetry,
should hold.

In this short paper the question of calculating the heavy neutrino mass scale
from experimental data and quark-lepton symmetry is addressed
\cite{calc1,calc2,calc3},
assuming $|m_3| \gg |m_{1,2}|$, where $m_i$ are the light neutrino masses,
and it is shown that such a scale is at or above the unification scale,
unless $m_1 \simeq -m_2$. For nearly opposite $m_1$, $m_2$, the heavy neutrino
mass scale is around the intermediate scale of a GUT such as $SO(10)$
with $SU(4) \ts SU(2) \ts SU(2)$ as intermediate symmetry \cite{10}.

Assuming maximal mixing for atmospheric neutrinos
and that the electron neutrino does not contribute to the atmospheric
oscillations \cite{max},
the matrix $M_L$ can be written as \cite{alt}
\ini
M_L= \left( \begin{array}{ccc}
            2 \epsilon & \delta & \delta \\
            \delta & \sigma & \rho \\
            \delta & \rho & \sigma
           \end{array} \right)
\fin
with
\inir
\epsilon &=& \frac{1}{2}(m_1 c^2+ m_2 s^2) \\
\delta &=& \frac{1}{\sqrt2}(m_1-m_2)cs \\
\sigma &=& \epsilon_2 + \frac{m_3}{2} \\
\rho &=& \epsilon_2-\frac{m_3}{2} \\
\epsilon_2 &=& \frac{1}{2}(m_1 s^2+ m_2 c^2),
\finr
and $c=\cos \theta$, $s=\sin \theta$,
where $\theta$ is the mixing angle of solar
neutrinos. The inverse of $M_L$
is given by
\ini
M_L^{-1}= \left( \begin{array}{ccc}
            \sigma^2-\rho^2 & \delta(\rho-\sigma) & \delta(\rho-\sigma) \\
            \delta(\rho-\sigma) & 2\epsilon\sigma-\delta^2 &
             \delta^2-2\epsilon\rho \\
            \delta(\rho-\sigma) & \delta^2-2\epsilon\rho &
            2\epsilon\sigma-\delta^2
           \end{array} \right)
\frac{1}{m_1 m_2 m_3}.
\fin        
Without loss of generality we can assume $m_3 >0$. Moreover, if
$m_3 \gg |m_2|,~|m_1|$, with $\Delta m^2_{sun}=m_2^2-m_1^2$ and
$\Delta m^2_{atm}=m_3^2-m_{1,2}^2$, then
\ini
M_L^{-1}= \left( \begin{array}{ccc}
            0 & -\delta & -\delta \\
            -\delta & \epsilon & \epsilon \\
            -\delta & \epsilon & \epsilon
           \end{array} \right)
\frac{1}{m_1 m_2}.
\fin
In the basis where the charged lepton mass matrix $M_e$ is diagonal,
for $M_D$ we take
\ini
M_D=\frac{m_{\tau}}{m_b}~ $diag$ (m_u,m_c,m_t).
\fin
An almost diagonal form of this kind for $M_D$ is again suggested by
quark-lepton symmetry and GUTs \cite{rrr},
and the factor is due to running from unification or intermediate scale
\cite{run}, corresponding to supersymmetric and nonsupersymmetric cases,
respectively. From the seesaw formula (1) and eqn.(10) we obtain
\ini
M_R=M_D M_L^{-1} M_D
\fin
and, due to the form of $M_L^{-1}$ and $M_D$ in eqns.(9),(10), 
two elements of $M_R$ have to be considered to discover
the scale of heavy neutrino mass:
\inir
M_{R33}&=&k~ \frac{m_1 c^2+m_2 s^2}{m_1 m_2}~m_t^2, \\
M_{R13}&=&k~ \frac{1}{\sqrt2} \frac{(m_1 -m_2 )cs}{m_1 m_2}~m_u m_t, 
\finr
with $k=(m_{\tau}/m_b)^2$. We will consider two cases for $s$, namely
$s=0$ (single maximal mixing) and $s=1/\sqrt2$ (bimaximal mixing \cite{bim}).
For $s=0$ we get the scale
\ini
M_{R33}=k~ \frac{m_t^2}{m_2}.
\fin
For $s=1/\sqrt2$ we have three subcases: $|m_2| \gg |m_1|$ which gives
\ini
M_{R33}=k~ \frac{1}{2} \frac{m_t^2}{m_1};
\fin
$m_2 \simeq m_1$ yielding
\ini
M_{R33}=k~  \frac{m_t^2}{m_{1,2}};
\fin
and $m_2 \simeq -m_1$ for which the scale is given by  
\ini
M_{R13}=k~ \frac{1}{\sqrt2} \frac{m_u m_t}{m_{1,2}},
\fin
while $M_{R33}$ is much smaller.
The case $s=0$ is near the small mixing MSW
solution of the solar neutrino problem \cite{bgg}, while the case $s=1/\sqrt2$
is near both the large mixing MSW and especially the vacuum oscillations
\cite{bgg}. 
Using the same numerical values of ref.\cite{calc2}
we find $M_{R33} \gtrsim 10^{15}$ GeV for $s=0$.
For $s=1/\sqrt2$, the three eqns.(15),(16) and (17) lead respectively to: 
$M_{R33} \gtrsim 10^{16}$ GeV (large mixing MSW), $M_{R33} \gtrsim 10^{18}$ GeV
(vacuum oscillations); $M_{R33} \gtrsim 10^{15}$ GeV (LMSW and VO);
$M_{R13} \gtrsim 10^{10}$ GeV (LMSW and VO).
Therefore we find that the heavy neutrino mass scale can be at the
intermediate scale only if $m_2 \simeq -m_1$ (see ref.\cite{come}),
and $M_R$ has a roughly off-diagonal form.
As the intermediate scale is related to the nonsupersymmetric case \cite{dk},
we get that nearly opposite masses are also related to that case
(we are in a CP conserving framework, the mass sign corresponds to CP parity
\cite{bgg}).
For positive masses the two MSW
solutions are in agreement with the unification scale while the vacuum
oscillation solution gives a scale well above the unification scale, unless
$m_2 \simeq m_1$.
It is worth noting that a smaller mass is obtained for opposite $m_1$, $m_2$,
due to the appearance of $m_u m_t$, rather than $m_t^2$, in eqn.(17). 
Moreover, from eqns.(14-17) one can read the dependence of the
heavy neutrino mass scale
on light neutrino masses. This dependence is in agreement with the
analyses given in refs.\cite{calc2,calc3}, where only positive values of
the light neutrino masses are considered.
The three eigenvalues of $M_R$ have a strong hierarchy, unless there is one
negative mass, in which case the structure of $M_R$ is very different from
that of $M_D$.
Finally, we notice that the bimaximal case with opposite masses has a different
impact on the neutrinoless double-beta decay parameter
${\cal M}_{ee}=2 \epsilon$, with respect to the bimaximal case with positive
masses and that, for a hierarchical spectrum, neutrinos are not relevant
for hot dark matter.

In conclusion, we have calculated the scale of heavy neutrino mass in the
seesaw mechanism assuming hierarchical light neutrino masses, maximal
or bimaximal mixing (as suggested by experimental data) and quark-lepton
symmetry. The main results are: intermediate scale is obtained only if
the two lightest left-handed neutrinos have nearly opposite masses; the vacuum
oscillation case with positive and full hierarchical masses
leads to a scale near the Planck mass.

\end{document}